\def\nh{{n_{\rm H}}}
\def\nh2{{n(\rm H_2)}}
\def\h2{${\rm H_2}$}

\def\3cm{\rm {cm^{-3}}}
\def\2cm{\rm {cm^{-2}}}
\def\s-1{\rm {s^{-1}}}

\def\mum {\hbox{$\mu$m}}
\def\kms {\hbox{${\rm km\,s}^{-1}$}}
\def\Kkms {\hbox{${\rm K}\,{\rm km}\,{\rm s}^{-1}$}}

\def\twco{\hbox{$^{12}$CO}}

\def\ci{\hbox{\rm {[C {\scriptsize I}]}}}
\def\cii{\hbox{\rm {[C {\scriptsize II}]}}}
\def\hi{\hbox{\rm {H {\scriptsize I}}}}
\def\hii{\hbox{\rm {H {\scriptsize II}}}}
\def\c18o{\hbox{C$^{18}$O}}

\def\ciup{\hbox{\rm {[C {\scriptsize I}]}~$^3P_2-{^3P_1}$}}

\documentclass[letter]{aa} % for the letters
\usepackage{multirow}
\usepackage{graphicx}
\usepackage{epsfig}
\usepackage{subfig}
\usepackage[]{natbib}
\usepackage{txfonts}
\usepackage{psfrag}
\usepackage{units}
\begin{document}
\title{The ionized and hot gas in M17~SW:}
\subtitle{SOFIA/GREAT THz observations of [C II] and \twco~$J$=13--12}
\author{J.P.~P\'{e}rez-Beaupuits\inst{1} \and
        H.~Wiesemeyer\inst{1} \and        
        V.~Ossenkopf\inst{2} \and
        J.~Stutzki\inst{2} \and
        R.~G\"usten\inst{1} \and
        R.~Simon\inst{2} \and
        H.-W.~ H\"ubers\inst{3,4} \and
        O.~Ricken\inst{2,1} %\and
        %G.~Sandell\inst{5}
}
\offprints{J.P.~P\'erez-Beaupuits}
\institute{
 Max-Planck-Institut f\"ur Radioastronomie, Auf dem H\"ugel 69, 53121 Bonn, Germany
 \email{jp@mpifr.mpg.de} 
 \and
 I. Physikalisches Institut der Universit\"at zu K\"oln, Z\"ulpicher Stra\ss e 77, 50937 K\"oln, Germany
 \and
 Deutsches Zentrum f\"ur Luft- und Raumfahrt, Institut f\"ur Planetenforschung, Rutherfordstrasse 2, 12489 Berlin, Germany
 \and
 Institut f\"ur Optik und Atomare Physik, Technische Universit\"at Berlin, Hardenbergstraße 36, 10623 Berlin, Germany
 %\and
 %SOFIA-USRA, NASA Ames Research Center, Mail Stop N211-3, Building N211/Rm. 249, Moffett Field, CA 94035, USA
}
\date{Received 1 February 2012 / Accepted 6 March 2012}
\titlerunning{\cii\ and \twco\ $J$=13--12 in M17~SW}
%\authorrunning{ }

%
\abstract
  {} %leave it empty if necessary
  {With new THz maps that cover an area of $\sim$3.3$\times$2.1~pc$^2$ we probe the spatial distribution and association of the ionized, neutral and molecular gas components in the M17~SW nebula.
  }
  {We used the dual band receiver GREAT on board the SOFIA airborne telescope to obtain a 5$'$.7$\times$3$'$.7
  map of the \twco~$J$=13--12 transition and the \cii~158~\mum\ fine-structure line in M17~SW
  and compare the spectroscopically resolved maps with corresponding ground-based data for low- and mid-$J$ CO and \ci\ emission.
  }
  {For the first time SOFIA/GREAT allow us to compare velocity-resolved \cii\ emission maps with molecular tracers.
   We see a large part of the \cii\ emission, both spatially and in velocity, that is completely non-associated with the other tracers of photon-dominated regions (PDR).
   Only particular narrow channel maps of the velocity-resolved \cii\ spectra show a correlation between the different gas components, which is not seen at all in the integrated intensity maps. These show different morphology in all lines but give hardly any information on the origin of the emission.
   The \cii~158~\mum\ emission extends for more than 2 pc into the M17~SW molecular cloud and its line profile covers a broader velocity range than the \twco~$J$=13--12 and \ci\ emissions, which we interpret as several clumps and layers of ionized carbon gas within the telescope beam.
   The high-$J$ \twco\ emission emerges from a dense region between the ionized and neutral carbon emissions, 
   indicating the presence of high-density clumps that allow the fast formation of hot CO in the irradiated
   complex structure of M17~SW. The \cii\ observations in the southern PDR cannot be explained with stratified nor clumpy PDR models.
  }
  {}

\keywords{galactic: ISM
--- galactic: individual: M17 SW
--- radio lines: ISM
--- molecules: \twco
--- atoms: \cii}

\maketitle

\section{Introduction}

M17~SW is a giant molecular cloud at a distance of $\sim$1.98 kpc \citep{xu11}, illuminated by a highly obscured ($A_v>10$ mag) cluster of several (among $\ga100$ stars) O/B stars \citep{beetz76, hanson97}, and it harbors several candidate young stellar objects \citep{povich09}. The large amount of observational data available in the literature and its nearly edge-on geometry make M17~SW one of the best-studied prototypes of a clumpy photon-dominated region PDR interface in the Galaxy. Studies of molecular and atomic emission indicate that the structure of the gas is highly clumped and not homogeneous \citep[][and references therein]{stutzki88, stutzki90, meixner92, pb10}, and the structure of its neutral and molecular gas seems to be dominated by magnetic rather than by thermal gas pressure, in contrast to many other PDR regions \citep{pellegrini07}.
Temperatures of $\sim$275 K were found toward the VLA 21cm continuum arc \citep{brogan01}, and are associated with NH$_3$ and highly excited \twco\ emission \citep{gusten88, harris87}.

Recent AKARI observations \citep{okada10} with spatial resolutions between 39$''$ and 57$''$ showed that the \cii~158~\mum\ line emission is widespread in the M17 complex and peaks at the northern (M17~N) and southern (M17~SW) bars, in agreement with previous observations \citep{matsuhara89, stutzki88}.
However, this and previous observations of the hot and ionized gas in M17~SW are limited in spatial resolution and extent \citep[e.g.][]{harris87, stutzki88, genzel88, meixner92, howe00}. Therefore, in this work we present new observations (of an area 5$'$.7$\times$3$'$.7) of hot molecular (\twco~$J$=13--12) and ionized atomic (\cii~158~\mum) gas, with spatial resolutions of $\sim$19$''$.8 and $\sim$15$''$.6, respectively, which advances existing work in M17~SW.

%----------------------------------------------------------------------------
\begin{figure*}[!tp]

\subfloat[]{\label{fig:maps-a}
\hspace{-0.6cm}
 \begin{minipage}[c]{.45\textwidth}
  \centering
  \vspace{-0.1cm}
  \begin{tabular}{l}
    \epsfig{file=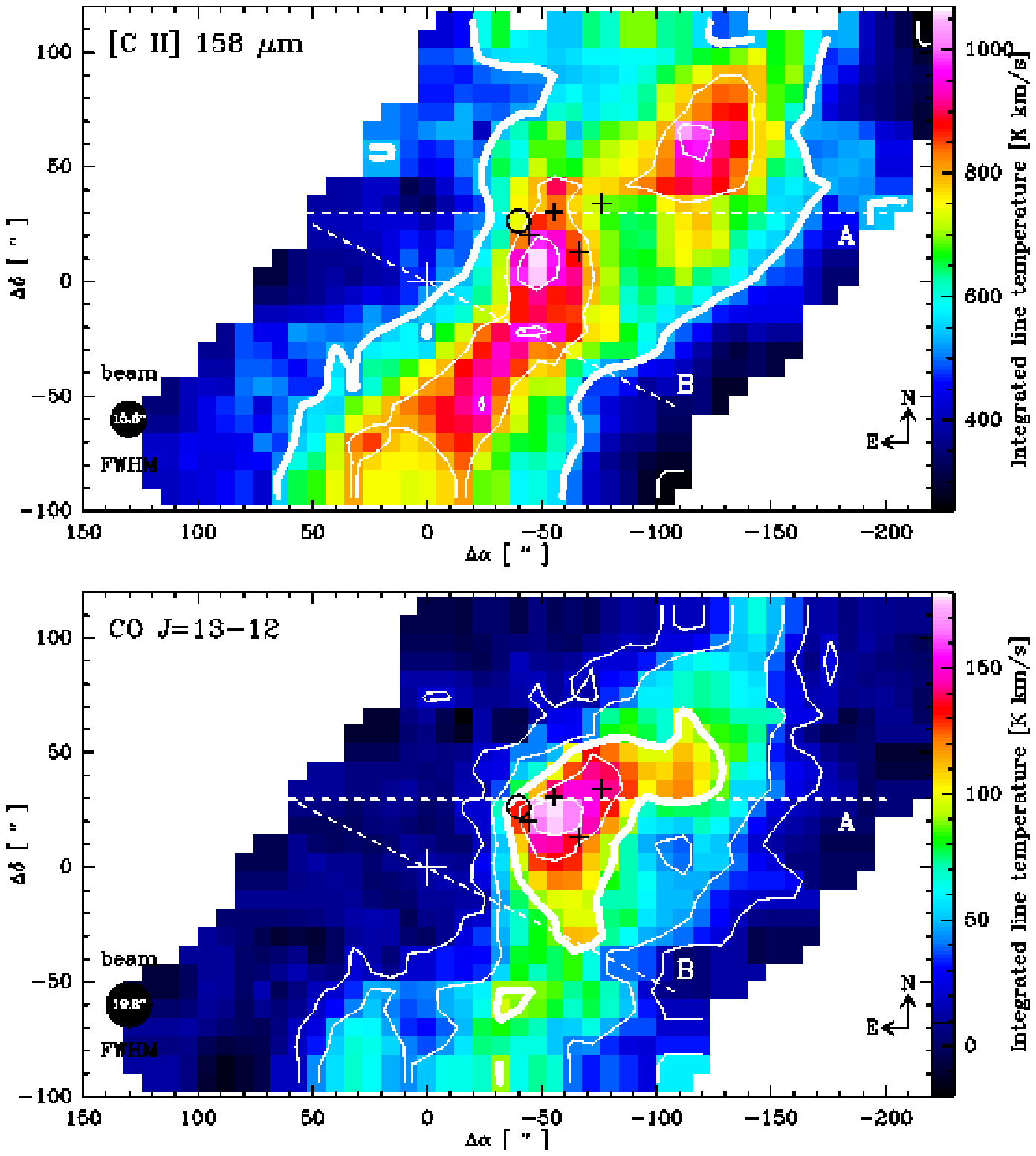,angle=0,width=0.85\linewidth}
  \end{tabular}
 \end{minipage}} %&
 \hspace{-0.0cm}
 \subfloat[]{\label{fig:maps-b}
 \begin{minipage}[c]{.49\textwidth}
  \vspace{-0.4cm}
  \centering
  \begin{tabular}{r}
   \epsfig{file=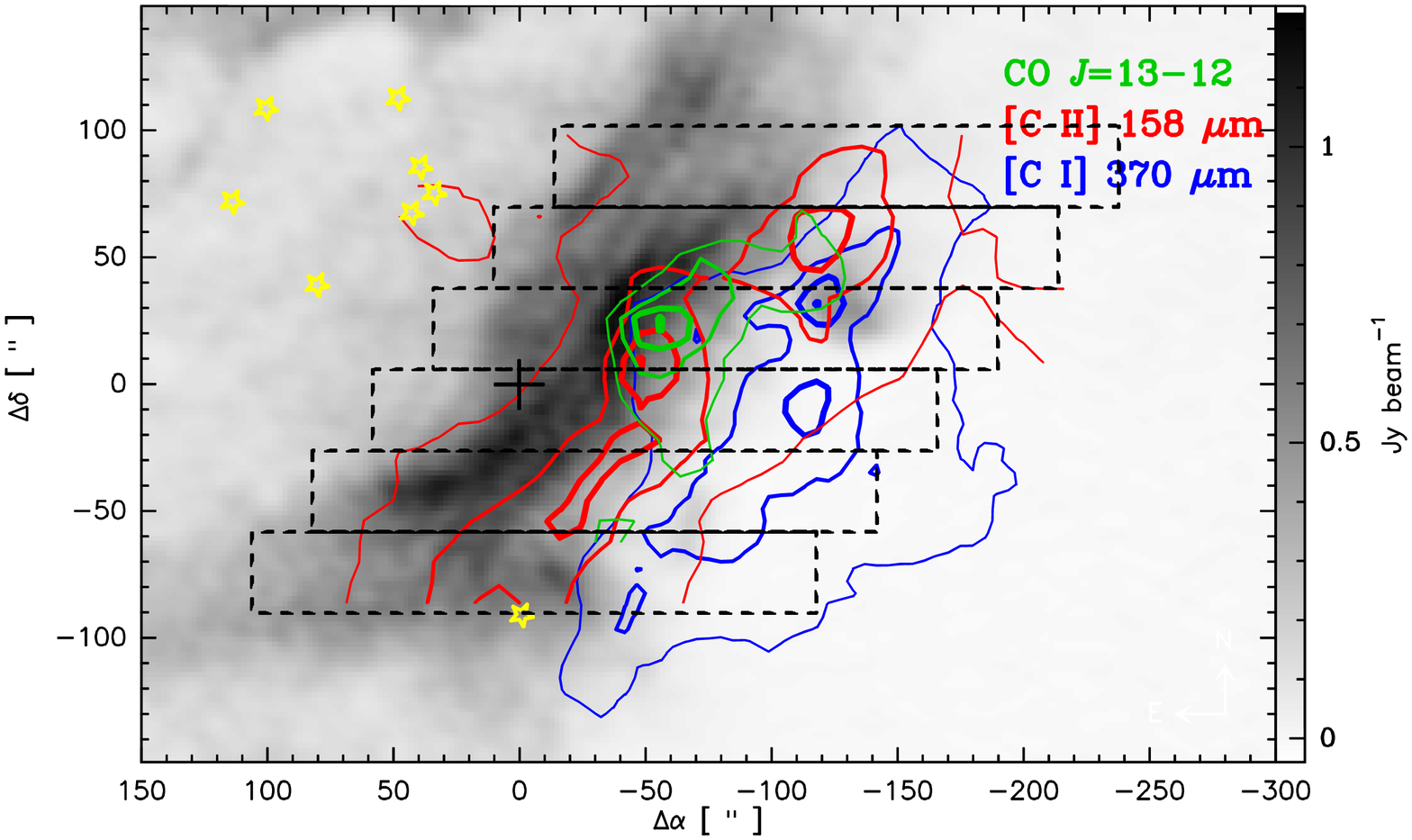,angle=-90,width=0.8\linewidth}
  \end{tabular}
 \end{minipage}}\\

\vspace{-3.6cm}
\hfill\begin{minipage}[r]{.59\textwidth}
 
    \caption{\footnotesize{\textit{Left panels} - Intensity maps of \cii~158 \mum\ (\textit{top}) and \twco\ $J$=13--12 (\textit{bottom}) in M17~SW, integrated in the velocity ranges 0--40~\kms\ and 14--28~\kms, respectively. The contour levels are the 10\%, 25\%, 50\% (thick line), 75\% and 90\% of the peak emission. Dashed lines are the strip lines shown in Fig.~\ref{fig:strip-lines}. The central position (0$''$,0$''$) is marked with a cross. The ultracompact \hii\ region M17-UC1 and four H$_2$O masers \citep{johnson98} are marked by the black circle and plus symbols, respectively. \textit{Right panel} - 21 cm continuum emission by \citet{brogan01} with the overlaid contours of the velocity integrated (same as above) emission of \twco~$J$=13--12 (\textit{green}), \cii\ (\textit{red}), and the \ciup~370~\mum\ (\textit{blue}, integrated in 9--30~\kms) from \citet{pb10}. The contour levels (from thin to thick) are the 50\%, 75\% and 90\% of the peak emission. The \textit{stars} indicate the O and B ionizing stars \citep{beetz76, hanson97}. Dashed frames depict the beam center for the edges of the 6 OTF strips. Contour maps are smoothed to $20''$ resolution.}}
   \label{fig:maps}

\end{minipage}

\end{figure*}
%----------------------------------------------------------------------------

%----------------------------------------------------------------------------
\begin{figure*}[!tp]
\vspace{-0.7cm}
 \begin{minipage}[l]{.55\textwidth}
  \begin{tabular}{@{\extracolsep{\fill}} ll}
   \epsfig{file=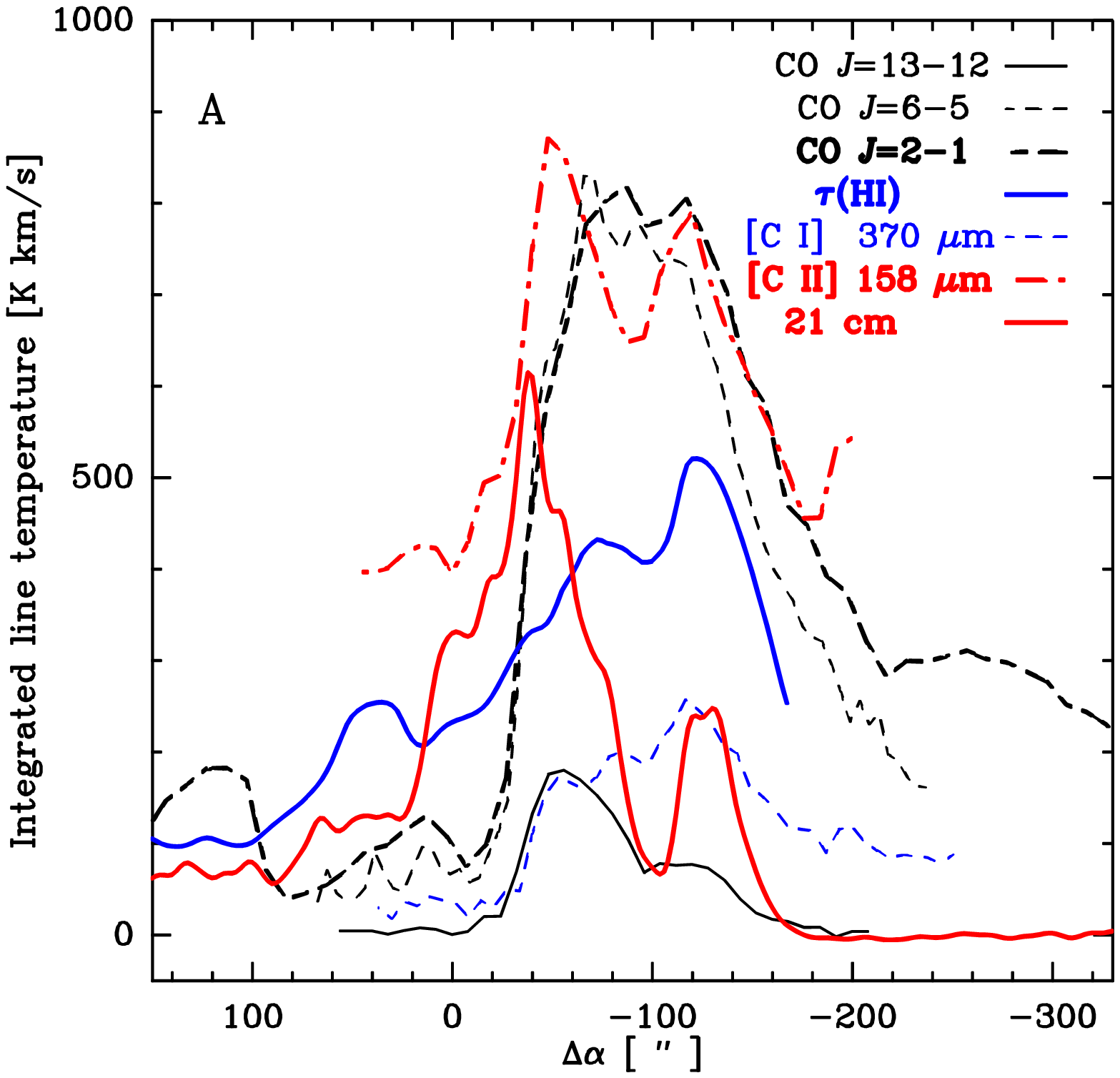,angle=-90,width=0.48\linewidth} &
   \hspace{-0.3cm}\epsfig{file=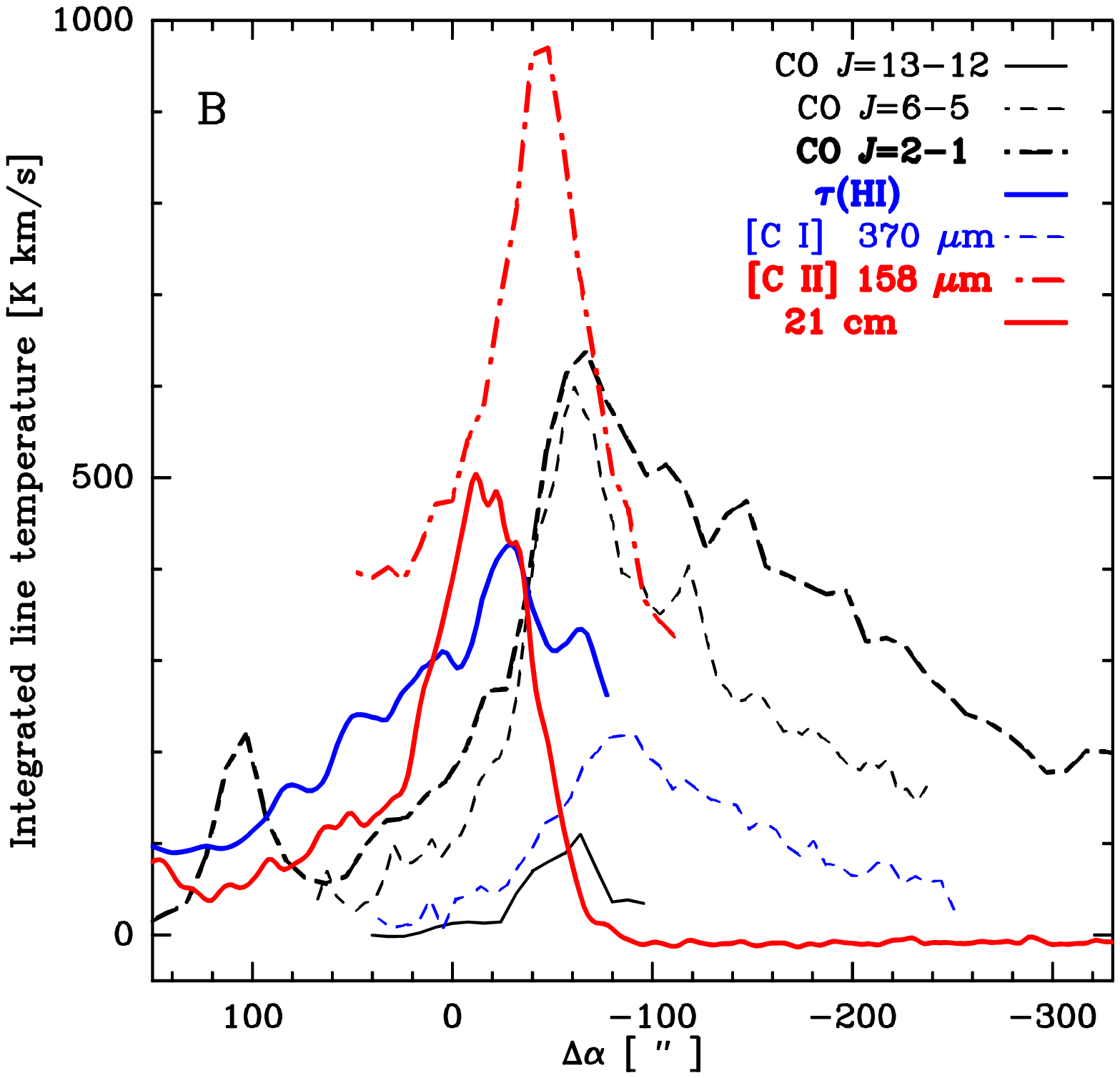,angle=-90,width=0.48\linewidth}
  \end{tabular}
 \end{minipage}
 \begin{minipage}[l]{.45\textwidth}
   \caption{\footnotesize{\textit{Left} - Strip lines of the molecular (\twco) and atomic (\ci, \cii) gas at $\Delta\delta=30''$ (P.A. $90^{\circ}$) across the ionization front of M17 SW. \textit{Right} - Strip lines at P.A. $63^{\circ}$ ($\Delta\delta=\Delta\alpha/2$). The lower-$J$ lines of \twco, and \ci\ are from \citet[][their Fig.~3]{pb10}. All these profiles are in units of \Kkms. The VLA 21~cm continuum and \hi\ optical depth (integrated between 0 and 30~\kms) by \citet{brogan01} are in units of 500$\times$Jy beam$^{-1}$ and 8$\times$$\tau({\rm HI})$ \kms, respectively. The offset, $\Delta\alpha=0''$ in R.A., is the same as in Fig.~\ref{fig:maps}.}}
   \label{fig:strip-lines}
 \end{minipage}

\end{figure*}
%----------------------------------------------------------------------------

\section{Observations}

The observations were performed with the German Receiver for Astronomy at Terahertz Frequencies \citep[GREAT\footnote{GREAT is a development by the MPI f\"ur Radioastronomie and the KOSMA/ Universit\"at zu K\"oln, in cooperation with the MPI f\"ur Sonnensystemforschung and the DLR Institut f\"ur
Planetenforschung.},][]{heyminck12} on board the Stratospheric Observatory For Infrared Astronomy (SOFIA).
We used the dual-color spectrometer during its first Short Science flight on 05 April 2011
to simultaneously map the fine-structure transition of \cii\ at 1900.536900~GHz (157.7~\mum) and the \twco~$J$=13--12 transition at 1496.922909~GHz (200.3~\mum) toward M17~SW.
The observations were performed in on-the-fly (OTF) total power mode. The area mapped consists of six strips, each covering $224''\times32''$ ($\Delta\alpha \times \Delta\delta$ with a sampling of $8''$ (half the beamwidth at 1.9 THz).
Hence, each strip consists of four OTF lines containing 28 points each. We integrated 1s per dump and 5s for the off-source reference.

All our maps are centered on R.A(J2000)=18:20:27.6 and Dec(J2000)=-16:12:00.9, which corresponds to the SAO star 161357. For better system stability we used a nearby reference position at offset (345$''$,$-230''$). A pointed observation of this reference position against the known better reference \citep[offset (1040$''$,$-535''$)][]{matsuhara89} showed that the reference is clean of \twco\ emission, but contains weak ($<$20\% of the peak emission) and relatively narrow (FWHM$\sim$15~\kms) \cii\ emission. All \cii\ spectra presented here were corrected for this missing flux.

Pointing was established with the SOFIA optical guide cameras, and was accurate to better than $10''$.
Because the acousto-optical and fast Fourier transform spectrometers \citep{klein12} operated in parallel give redundant information, we base the following analysis on the data from the latter, which provided %
1.5 GHz bandwidth with about 212 kHz ($\sim$0.03~\kms) of spectral resolution.
Because during this commissioning flight the instrument showed some random gain variations, we discarded 15-20\% of the spectra by filtering the spectra for outliers in the total-power IF level and for outliers in the noise rms (obtained after subtracting a third-order polynomial), retaining only reliable data.
The calibration of these data to antenna temperature was performed with the 
\textit{kalibrate} task from the \textit{kosma\_software} package \citep{guan12}. 
We then reduced and imaged the 
data further with the CLASS90 packages, which is part of the GILDAS\footnote{http://www.iram.fr/IRAMFR/GILDAS} software.
Using the beam efficiencies ($\eta_c$) 0.51 for \cii\ and 0.54 for \twco~$J$=13--12, and the forward efficiency ($\eta_f$) of 0.95 \citep{heyminck12},
we converted all data to main beam brightness temperature scale, $T_{B}=\eta_{f}\times T_{A}^{*}/\eta_{c}$.

\section{Results}

The velocity-resolved spectra of the new GREAT/SOFIA observations reveal a much more complex structure than expected based on our understanding of M17~SW as a highly clumped PDR, with a significant fraction of the \cii\ emission not at all associated with the molecular PDR material. 

Figures 1--4 display the GREAT data compared with supplementary data of typical PDR tracers from the literature.
Figure~\ref{fig:maps} shows the velocity-integrated emission of \cii\ between 0 and 40~\kms\ (peak $\sim$1070 \Kkms) and \twco~$J$=13--12 [14 -- 28 \kms, peak $\sim$180 \Kkms] (left panels).
While the high-excitation CO emission follows the hot dense gas structure seen in previous submm line studies (but see the discussion below),
the \cii\ distribution reveals surprises:
its velocity-integrated distribution is quite shallow and far more extended than predicted for the stratification in classical PDR models. 
This is demonstrated in Fig.~\ref{fig:maps-a} and the intensity cuts across the PDR in Fig.~\ref{fig:strip-lines}, showing strong \cii\ emission over the whole extent of the mapped area. 

\begin{figure*}[!tp]
\vspace{-0.5cm}
 \begin{minipage}[l]{.70\textwidth}
\vspace{-0.3cm}
   \hspace{-0.15cm}\epsfig{file=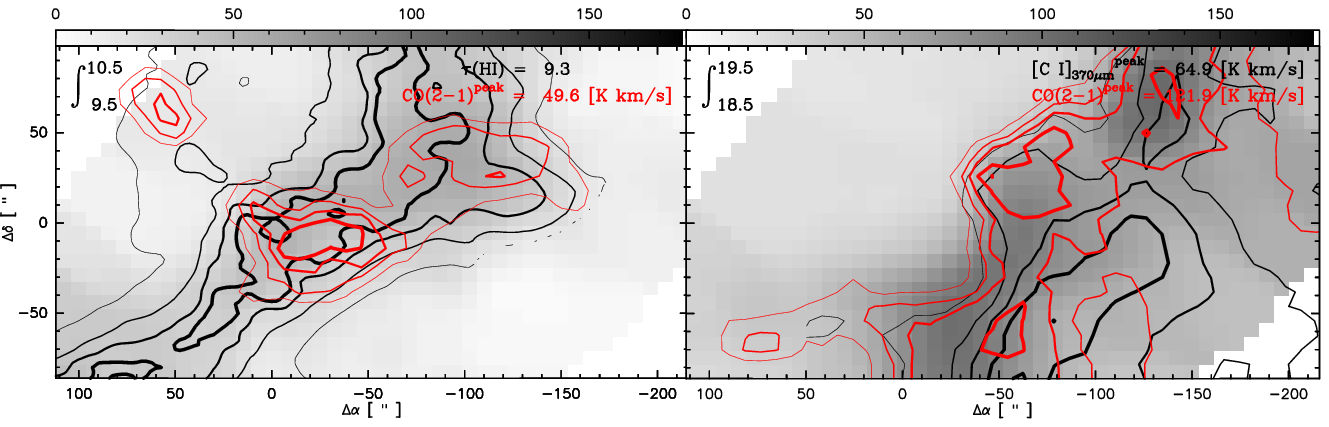,angle=0,width=0.965\linewidth}
 \end{minipage}
 \hspace{-0.5cm}
 \begin{minipage}[l]{.30\textwidth}
  \vspace{0.3cm}
  \caption{\footnotesize{Selected 1~\kms\ wide channel maps (gray scale) of the \cii\ velocity-integrated (\Kkms) in the velocity ranges 9.5--10.5~\kms\ and 18.5--19.5~\kms. The contours show the integrated emission in the corresponding velocity channels of \ci, \twco~$J$=2--1, and $\tau(\rm HI)$ in 20\% steps (from thin to thick lines) of the peak channel integrated values (top right in the maps). All maps but $\tau$(\hi) were smoothed to a resolution of 20$''$.}}
  \label{fig:CII-CI2-1-CO2-1-chan-map}
 \end{minipage}

\end{figure*}

The broader distribution is consistent with the lower angular resolution data from the KAO \citep{stutzki88}, at the time interpreted to result from the deeper UV-penetration into a clumpy medium. However, the new GREAT/SOFIA data, resolving the spectra in velocity, show a more complex scenario. 
We find that (in M17~SW) the \cii\ emission ($E_{u}$=91.21 K, and $n_{crit}\sim5\times10^3~\3cm$) traces - in addition to the dense PDR gas west of the ionization front - also a widespread (toward the east), probably more diffuse atomic gas component that is invisible in CO. A sample of this is shown in Fig.~\ref{fig:CII-CI2-1-CO2-1-chan-map} (left panel), displaying in gray scale the intensity distribution of the lower velocity components of \cii\ and its (non)correlation with CO and \hi.
Comparing individual spectra (Fig.~\ref{fig:spectra}), we see that the span of observed velocities in \cii\ is much wider than that of any other PDR tracing species. In the spectra at offset (0,0), a position well in front of the PDR interface (Fig.~\ref{fig:maps-b}), for example, \cii\ still shows a bright and amazingly wide line with $\Delta$V $\sim$ 30 \kms, including velocities \underline{not} detected in molecular (CO) or (denser) atomic (neutral carbon \ci) gas column density tracers. Most interestingly, those velocities (e.g., 4 -- 10 \kms) exhibit appreciable opacities in atomic hydrogen (\citealt{brogan01} - note that \hi\ is observed in absorption against the 21 cm background continuum, hence its detectability across the map depends on the continuum distribution). The broad and intricate structure of the \cii\ spectral line (Fig.~\ref{fig:spectra}) can also be the signature of several clumps or layers of ionized carbon gas at ``redder" and ``bluer" velocities than the other tracers.

Fig.~\ref{fig:maps-b} provides evidence for a C$^+\to$ C stratification across the PDR: the bulk of the \ci\ emission (defined as $>$75\% of the peak emission) avoids the bulk emission of \cii, with their peaks being separated by $\sim$70$''$ in the NE-SW direction, i.e., 0.67~pc at the distance of $\sim$1.98 kpc \citep{xu11}. 
In a previous analysis the extended \ci\ emission (west from the ionization front), and its particular peak emission in front of the bulk CO emission (if stratified, the \ci\ peak should be observed before the CO peak, from the direction of the PDR interface), was argued to emerge from the interclump regions of a very clumpy medium, but could also result from a partial face-on illumination of the molecular clouds \citep[e.g.,][]{stutzki88, stutzki90, meixner92, pb10}.
In the context of the clumpiness of the PDR and the newly discovered association of much of the \cii\ emission with atomic gas (see below) at velocities not matched by any molecular material, the \cii\ peak cannot be interpreted as the classical stratification expected in a homogeneous 
PDR\footnote{In a clumpy PDR, with many internal surfaces, the \cii\ peak should be even closer to the \ci\ peak, becoming indistinguishable from it depending on clumpiness of the source and spatial resolution of the observations.}, but instead has to be caused by excitation gradients with increasing optical depth into the atomic and molecular cloud.

The excitation study of the warm PDR layers, including the CO excitation, will be the subject of a forthcoming publication. Here we briefly elaborate on the new GREAT data:
The \twco~$J$=13--12 line follows a similar spatial distribution as the lower-$J$ lines reported in \citet{pb10}, although the whole \twco~$J$=13--12 emission is shifted toward the ionization front. Its peak emission is $\sim$16$''$ (0.15~pc) southeast of the \twco~$J$=6--5 peak. The $J$=13--12 line is already fainter than the lower-$J$ transitions, indicating subthermal excitation.

%---------------------------------------------------------------
%---------------------------------------------------------------

\section{Discussion}

%-------------------

We can distinguish two different spatial regimes in 
the PDR with embedded star cores of active star formation. 
The strip line along $\Delta\delta=30''$ (P.A. $90^{\circ}$, Fig.~\ref{fig:strip-lines}A) traces the embedded star-formation, going through the ultracompact \hii\ region UC1 ($\Delta\alpha=-30''$), the embedded H$_2$O maser positions farther west, and the weaker additional \hii\ region seen as a separate clump at $\Delta\alpha=-110''$ in the 21 cm continuum map and matching the secondary peak of \cii, \ci, and the \twco~$J$=2--1 lines (Figs.~\ref{fig:maps-b} and \ref{fig:strip-lines}A). A dominant internal heating is likely suppressing the layering here.
The strip with the position angle $63^{\circ}$ (Fig.~\ref{fig:strip-lines}B)
instead traces the classical PDR. For a homogeneous PDR one expects a
stratified layering of \cii\, \ci, high-$J$ CO and low-$J$ CO. However,
\citet{stutzki88} showed the structure to be very clumpy so that no
stratification should be observed. Our observations now show that high-$J$ CO and low-$J$ CO peak at similar depths in the
cloud while the integrated \cii\ peak is shifted by $\approx 0.19$~pc relative
to the CO peaks. The \cii\ shift cannot even be explained by homogeneous
PDR models, much less by a clumpy medium.

%-------------------

\begin{figure}[!tp]
 \vspace{-0.8cm}
 \centering

 \epsfig{file=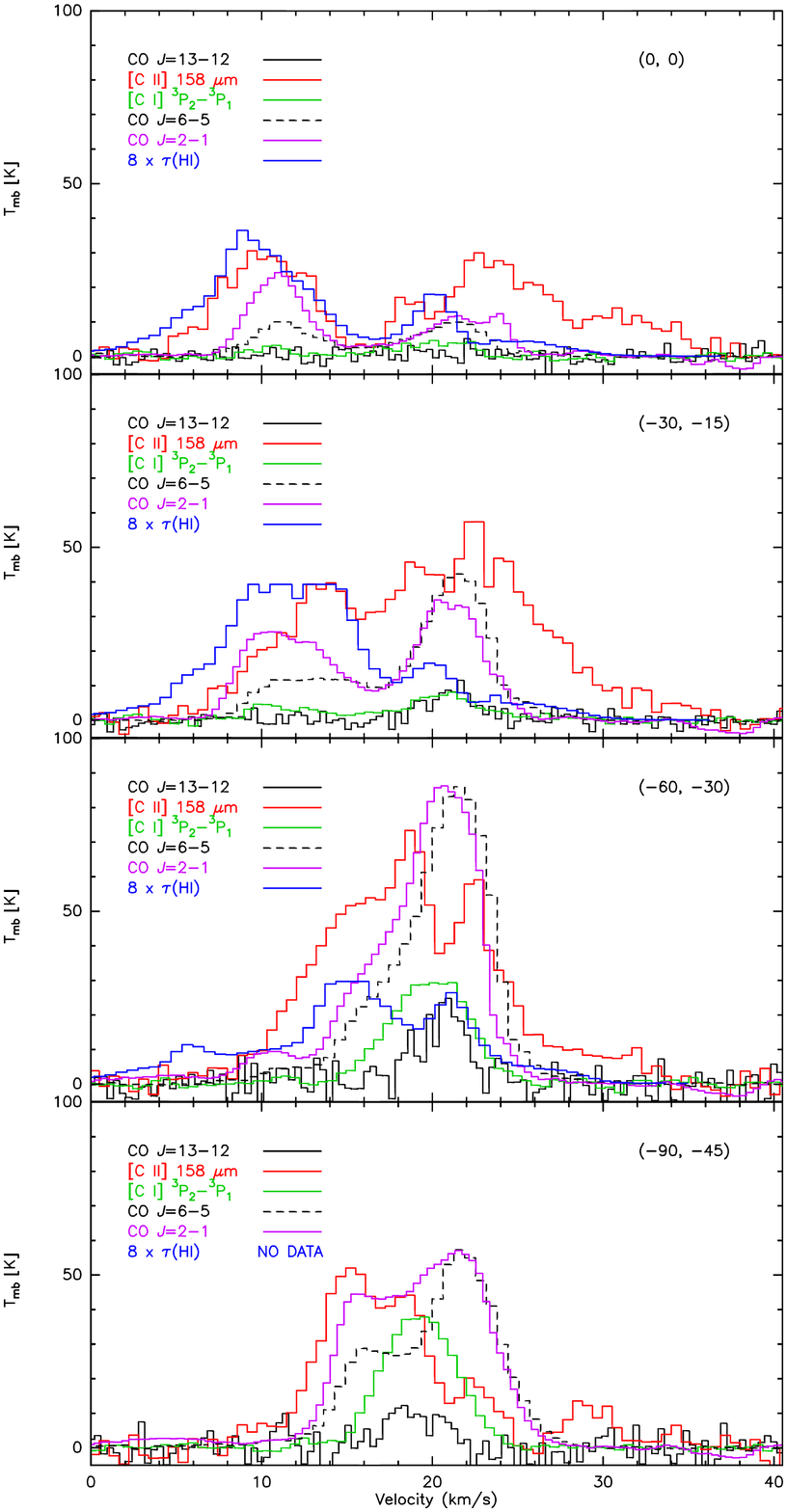,angle=0,width=0.73\linewidth}

 \vspace{-0.2cm}

 \caption{\footnotesize{Line profiles at selected positions along our cut across the PDR at P.A. $63^{\circ}$ ($\Delta\delta=\Delta\alpha/2$). All data but $\tau$(\hi) were smoothed to a spatial and a spectral resolution of 20$''$ and $\sim$0.6~\kms, respectively.}}

  \label{fig:spectra}
\end{figure}

In \cii\ we detect all velocity components that were discussed in 
the \hi\ channel maps by \citet[][their Figs. 4 and 5]{brogan01},
including a weak foreground cloud at 7~\kms{}, shocked gas in front
of the cluster that is blown toward the observer at 11--17~\kms{},
and the main PDR velocity of $\sim$21~\kms{}. In the higher density gas (west of the \hii\ region),
\cii\ seems to be optically thick, showing self-absorption dips
at the peak velocities of \hi\ and the molecular lines. The 
\twco~$J$=2--1 emission also traces the optically thin shocked gas down to 
10~\kms\ (Fig.~\ref{fig:spectra}), the \twco~$J$=13--12
line is hardly excited there.
A good match between \cii\ and \twco~$J$=2--1 can be seen at intermediate
velocities (Fig.~\ref{fig:CII-CI2-1-CO2-1-chan-map},
right panel) where the shifted \ci\ emission with respect to \cii\ is also evident.

We can compare the column density contributions visible in the
different velocity components using the column density 
for the \hi\ gas by \citet{brogan01} and the C$^+$ column density \citep[e.g.,][their eq.A.4]{schneider03}
responsible for the observed \cii{} emission. 
For the gas temperature and the density we assumed
250~K and $10^4~\3cm$ at the offset (0,0), and 150~K and $10^6~\3cm$ for 
the offsets deeper in the cloud according to \citet{pb10}.

To compare our results with the \hi\ data we estimated \cii\ and \hi\ column densities 
at selected positions along the southern PDR (Fig.~\ref{fig:spectra}) from the \cii\ emission
and \hi\ optical depths, integrating the spectra in the velocity ranges 11--17~\kms\
and 17--24.5~\kms. 
Selection of these velocity ranges is described by \citet{brogan99}.
In the range 11--17~\kms\ we found $N$(\cii) (with uncertainties of $\sim$20\%) between 6.5$\times$10$^{17}~\2cm$ at offset position (0,0) and 1.6$\times$10$^{18}~\2cm$ at ($-60$, $-30$), while the corresponding \hi\ column densities vary (with uncertainties 7--16\%) between 5.5$\times$10$^{21}~\2cm$ and 4.7$\times$10$^{21}~\2cm$, respectively. In the velocity range 17--24.5~\kms\ the \cii\ column densities are between 1.2$\times$10$^{18}~\2cm$ and 2.9$\times$10$^{18}~\2cm$ at the same offset positions, and we found $N$(\hi) between 3.7$\times$10$^{21}~\2cm$ and 4.4$\times$10$^{21}~\2cm$. The variation in HI column densities is less than 20\% in both velocity ranges, indicating a relatively homogeneous \hi\ interface seen also by \cii. The latter, instead, presents a much stronger column variation (40\%--60\%) depending on the ambient conditions.

Assuming the normal cosmic carbon abundance of $X(\rm C/H)=1.2\times10^{-4}$ \citep{wakelam08}
and complete ionization of the carbon, we obtained gas column densities $N({\rm H})=(5.4-13)\times
10^{21}~\2cm$ for the low-velocity component and $N({\rm H})=(10-24)\times 10^{21}~\2cm$
for the high-velocity component. The \cii\ emission of the low-velocity component in front of the PDR is
therefore consistent with pure \hi\ gas, while for all other components (above 17~\kms) at least half of the
\cii{} emission stems from molecular gas, i.e., classic PDR material.
The material in front of the southern PDR must be atomic to a high degree,
indicating non-equilibrium chemistry \citep[e.g.][]{stoerzer98}.

\section{Final remarks}

The integrated intensity maps do not tell us about the origin of the emission lines.
Only narrow channel maps of the velocity-resolved \cii\ spectra show a correlation 
with the diffuse gas components.
From the comparison of different tracers in different velocity components we found
that for the regions of embedded star formation the internal heating is always 
the dominant process, providing a very good match of all tracers without
significant layering.  The good spatial correlation of CO, \cii, and other 
tracers at particular velocity components/channels confirms the clumpy PDR picture from 
\citet{stutzki90}. 
However, a significant fraction of the \cii\ velocity components is not associated (spatially)
with the dense (high-$J$ CO) or the diffuse (\hi{} and CO $J$=2--1) PDR tracers.
The structure neither matches a stratified nor a clumpy PDR model. The \cii{}
emitting gas is only partially explained by atomic gas. Optically thin, hot molecular
gas must contribute more than half of the emission.

%---------------------------------------------------------------
%---------------------------------------------------------------

\acknowledgements{
We are grateful to C. Brogan for providing the 21 cm and \hi\ data.
We thank the referee for the very careful reading of the manuscript and constructive comments.
Results are partially based on observations made with the NASA/DLR Stratospheric Observatory
for Infrared Astronomy. SOFIA Science Mission Operations are conducted jointly by the
Universities Space Research Association, Inc., under NASA contract NAS2-97001,
and the Deutsches SOFIA Institut under DLR contract 50 OK 0901. 
We also gratefully acknowledge the outstanding support by the observatory staff.
This work was supported by 
the German \emph{Deut\-sche For\-schungs\-ge\-mein\-schaft, DFG\/} project SFB 956C.
}

\bibliographystyle{aa}
\setlength{\bibsep}{-2.1pt}
\bibliography{m17sw}

\end{document}